\documentclass[twocolumn,tighten,astrosymb,trackchanges]{aastex631}
\usepackage{multirow}
\let\tablenum\relax
\usepackage{siunitx}
\usepackage{amsmath,amsthm,amsfonts,amssymb,amscd}
\usepackage{booktabs}
\usepackage{tablefootnote}
\usepackage{apjfonts}

\definecolor{maroon}{rgb}{0.760,0.118,0.337}

\definecolor{darkaqua}{rgb}{0.0,0.45,0.65}

\def\cm{\mbox{\,cm}}
\def\cm3{\mbox{\,cm$^{-3}$}}

\newcommand{\LCO}{\affiliation{Las Cumbres Observatory, 6740 Cortona Drive, Suite 102, Goleta, CA 93117-5575, USA}}
\newcommand{\UCSB}{\affiliation{Department of Physics, University of California, Santa Barbara, CA 93106-9530, USA}}
\newcommand{\UCSC}{\affiliation{Department of Astronomy and Astrophysics, University of California, Santa Cruz, CA 95064-1077, USA}}
\newcommand{\CIERA}{\affiliation{Center for Interdisciplinary Exploration and Research in Astrophysics and Department of Physics and Astronomy, \\Northwestern University, 1800 Sherman Avenue, 8th Floor, Evanston, IL 60201, USA}}
\newcommand{\Rutgers}{\affiliation{Department of Physics and Astronomy, Rutgers, the State University of New Jersey,\\136 Frelinghuysen Road, Piscataway, NJ 08854-8019, USA}}

\newcommand{\ICG}{\affiliation{Institute of Cosmology \& Gravitation, University of Portsmouth, Dennis Sciama Building, Burnaby Road, Portsmouth PO1 3FX, UK}}
\newcommand{\AMNH}{\affiliation{Department of Astrophysics, American Museum of Natural History, Central Park West and 79th Street, New York, NY 10024-5192, USA}}

\shorttitle{\textit{HST} Late-Time Observations of SN~2012Z}
\shortauthors{Schwab et al.}
\submitjournal{\textit{The Astrophysical Journal}}
\received{31 March 2025}

\begin{document}

\title{The Remarkable Late-Time Flux Excess in Hubble Space Telescope Observations of the Type Iax Supernova~2012Z}

\correspondingauthor{Michaela Schwab}
\email{michaela.schwab@rutgers.edu}

\author[0009-0002-5096-1689]{Michaela Schwab}
\Rutgers
\author[0000-0003-3108-1328]{Lindsey A.\ Kwok}
\thanks{CIERA Postdoctoral Fellow}
\CIERA
\author[0000-0001-8738-6011]{Saurabh W.\ Jha}
\Rutgers
\author[0000-0001-5807-7893]{Curtis McCully}
\LCO
\UCSB
\author[0000-0002-4391-6137]{Or Graur}
\ICG
\AMNH
\author[0000-0002-2445-5275]{Ryan J.\ Foley}
\UCSC
\author[0000-0002-9830-3880]{Yssavo Camacho-Neves}
\Rutgers
\author[0000-0002-8092-2077]{Max J.\ B.\ Newman}
\Rutgers
\author[0000-0003-2037-4619]{Conor Larison}
\Rutgers
\author[0000-0001-8023-4912]{Huei Sears}
\Rutgers

\begin{abstract}
Type Iax supernovae (SNe~Iax) are thermonuclear explosions of white dwarfs, peculiar and underluminous compared to the normal type Ia supernovae. Observations of SNe~Iax provide insight into the physics of white-dwarf explosions and suggest that some may not be terminal events. Late-time photometry ($\sim$1400 days post-peak) of the type Iax SN~2012Z, the only white dwarf supernova with a pre-explosion detection of a progenitor system, revealed a flux excess that may be explained by a gravitationally bound remnant driving a radioactively powered wind. We present further late-time \emph{Hubble Space Telescope} photometry of SN~2012Z, $\sim$2500 days after the explosion, and find that the SN is still brighter than, but trending towards, the pre-explosion flux. Additionally, we observe that the excess F555W flux seen in previous data has grown more pronounced. The color of the excess flux disfavors a light echo or interaction with the circumstellar material. The decline rate of the excess flux is consistent with energy deposition from $^{55}$Fe, but the luminosity is higher than expected from models of the ejecta, further suggesting evidence for a bound remnant. Combining our data with future observations should allow for the detection of emission from the ejecta shock-heating of the companion helium star seen in the progenitor system.
\end{abstract}


\section{Introduction \label{sec:intro}} 
 White-dwarf supernovae are the thermonuclear explosions of white dwarfs (WD) in binary systems \citep{HoyleandFowler:1960,Wang:2012}. For ``normal'' Type Ia supernovae (SNe Ia), used as standardizable candles for cosmological measurements, the explosion is thought to fully disrupt the WD. Normal SNe~Ia comprise the majority of thermonuclear SNe, but growing populations of observed WD SNe deviate substantially photometrically and spectroscopically from the behavior that characterizes normal SNe Ia \citep{Taubenberger:2017, Jha:2019}. 

Type Iax supernovae (SNe~Iax), or 02cx-like SNe~Ia (named after the class prototype SN~2002cx; \citealt{Li:2003}), comprise the most populated class of peculiar WD SNe. SNe~Iax are estimated to occur at a rate of $15\%-30\%$ of the rate of normal SNe~Ia \citep{Foley:2013, Miller:2017, Perley:2020, Srivastav:2022}, and are peculiar and underluminous relative to normal SNe~Ia \citep{Li:2003, Jha:2006, Foley:2013, Jha:2017}. 

Unlike the relatively homogenous class of normal SNe~Ia, SNe~Iax exhibit a wide range of peak absolute magnitudes from nearly as bright as typical SNe Ia to 100 times fainter \citep{Jha:2017,Taubenberger:2017}. Like SNe~Ia, the peak luminosities, late-time photometry, and modeling of SNe~Iax suggest their light curve is powered by the radioactive decay of $^{56}$Ni $\rightarrow \, ^{56}$Co $\rightarrow \, ^{56}$Fe \citep{McCully:2014b}. Photometric analysis suggests that the lower luminosities observed in SNe~Iax are produced by lower masses of $^{56}$Ni synthesized in the explosion. 
 
At early times, SNe~Iax are spectroscopically similar to SNe~Ia, but while the ejecta of SNe~Ia cool and expand at late times, SNe~Iax continue to exhibit spectral features from an optically thick photosphere even at hundreds of days post-explosion \citep{Camacho-Neves:2023}. \cite{Foley:2016} attribute this lasting photosphere and the slow late-time decline of SNe Iax to contributions from not only the SN ejecta, but also a radioactive wind driven by a bound remnant.

The leading explosion model for a SN~Iax is a weak deflagration (the burning front propagates subsonically), consistent with the lower ejecta velocities and relatively small amount of radioactive nickel observed in SNe~Iax \citep[e.g.,][]{Foley:2013,Camacho-Neves:2023}. In this model, the progenitor WD may not be completely disrupted \citep{Magee:2016}, instead leaving behind a gravitationally bound remnant polluted with ashes of nuclear burning \citep{Fink:2014}. The decay of these radioactive burning products in the bound remnant can power a wind. Trapping of the decay energy in relatively high-density material could lead to observable emission at epochs when the original ejecta have otherwise expanded and diluted (allowing most of the decay energy in the ejecta to escape, e.g., in gamma-rays rather than optical light). \cite{ShenSchwab:2017} found that the flux of the remnant should dominate that of the ejecta a few years after the explosion. \cite{Vennes:2017} and \cite{Raddi:2019} show evidence for the direct detection of a potential SN~Iax bound remnant in the Milky Way, and \cite{Foley:2014} suggest a bound remnant to explain a coincident point source to the extremely faint SN~Iax~2008ha observed four years after explosion. 

SN Iax 2012Z is the only WD SN (including normal SNe Ia) for which a progenitor system has been detected in pre-explosion imaging \citep{McCully:2014, McCully:2022}. SN~2012Z exploded as a luminous SN Iax in NGC~1309, a galaxy for which, serendipitously, deep \emph{Hubble Space Telescope} (\textit{HST}) imaging had already been obtained for the discovery and monitoring of Cepheid variable stars \citep{Riess:2009,Riess:2009b}. This host-galaxy imaging revealed a luminous blue progenitor system, interpreted to be a helium star donor to the progenitor white dwarf \citep{McCully:2014}. 

Late-time photometry of SN~2012Z by \cite{McCully:2022} showed a flux excess in the light curve, compared to expectations based on the earlier decline rate or from radioactive decay, beginning around 500~days after maximum and persisting through observations around $+$1400~days. Possible interpretations of this excess flux included a shock-heated companion or interaction with the surrounding circumstellar medium (CSM), but emission from a bound remnant at the location of SN~2012Z was a promising explanation.

Here we present further late-time photometry of SN~2012Z at $\sim$2500 days post maximum brightness. In \autoref{sec:Obs} we describe our observations along with the existing pre-explosion and late-time data \citep{McCully:2014,McCully:2022} included in our analysis. In \autoref{sec:analysis} we analyze the late-time light curve and spectral energy distribution (SED), and compare to bound remnant wind and shock-heated companion models. We discuss the implications of our latest data for constraining explosion scenarios for SN Iax in \autoref{sec:conclusions}.

\begin{figure*}
    \centering
    \includegraphics[width=0.8\textwidth]{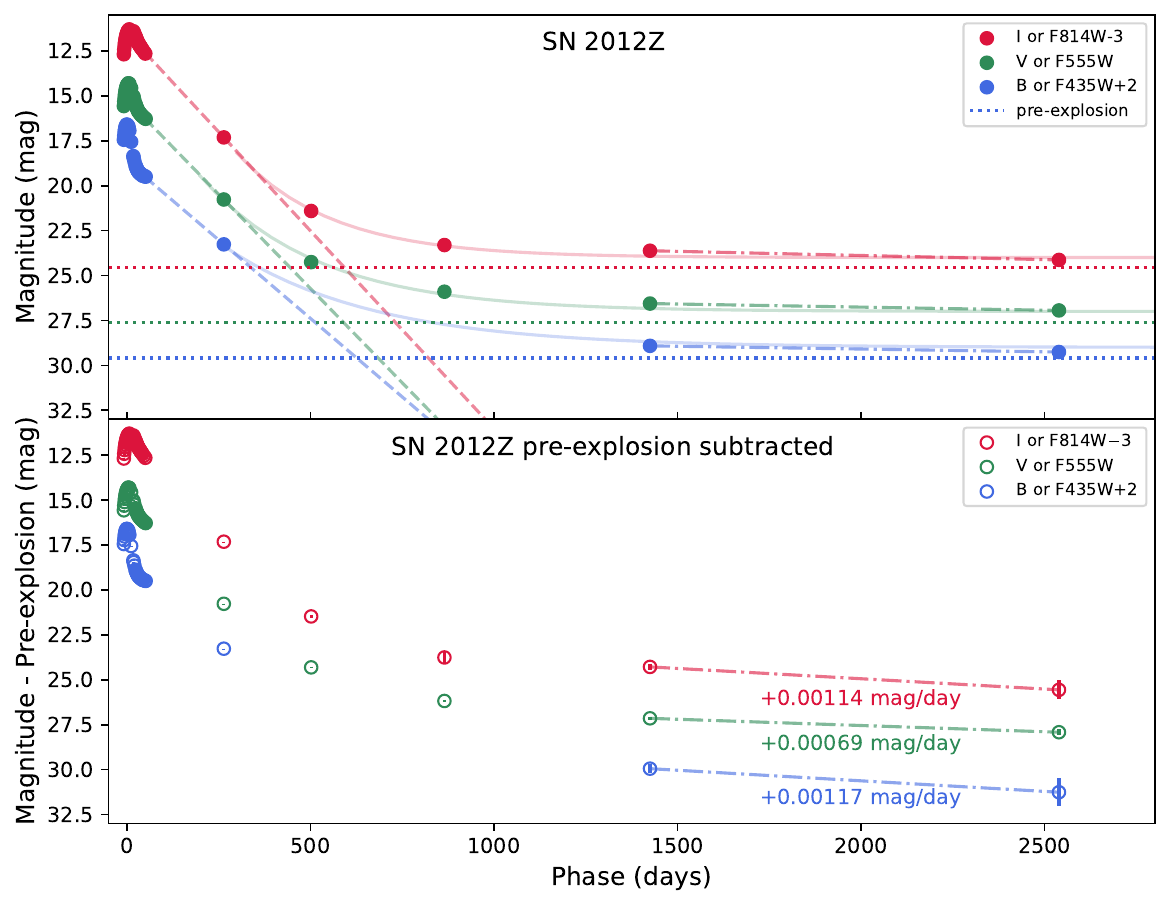}
    \caption{Ground-based (early-time) and \textit{HST} (after +200 days) photometry of SN~2012Z extending to $\sim2500$ days post-maximum (with the latest observation from 2019). Dashed lines in the top panel show the predicted decay of the supernova ejecta flux (extrapolated from earlier data) that should play a negligible role at the later epochs. The dotted lines show the pre-explosion flux \citep{McCully:2014}. The observations asymptote towards the pre-explosion flux (solid lines) suggesting that the progenitor system is still present, in addition to an unexplained excess flux. After subtraction of the pre-explosion flux (lower panel), we see the excess flux is also declining with time but is most prominent and slowest declining in F555W (green).}
    \label{fig:light-curve}
\end{figure*}

\section{Observations \label{sec:Obs}}

We present late-time \emph{HST} observations of SN~2012Z at $+$2539 days post-maximum from program GO-15205 (PI: Jha) taken with the Advanced Camera for Surveys Wide Field Channel (ACS/WFC) in the F435W, F555W, and F814W bandpasses (corresponding roughly to $B$, $V$, and $I$, respectively). These data were taken approximately 1100 days after the last published epoch \citep[from 2019;][]{McCully:2022} that was also obtained with ACS/WFC in the same filters. We compare this most recent photometry with the pre-explosion \citep{McCully:2014} and post-explosion \emph{HST} imaging \citep{McCully:2022}, taken with ACS/WFC as well as the Wide Field Camera 3 (WFC3) with UVIS.

Data processing of the \emph{HST} ACS/WFC images began with pipeline-calibrated FLC images, as described by \cite{2021wfcd.book....5S}, with the additional step of pixel-based charge transfer efficiency (CTE) correction \citep{AndersonandBendin:2010:}. We aligned all images using TweakReg in DrizzlePac \citep{2012drzp.book.....G} and measured point-spread function (PSF) photometry using DOLPHOT \citep{Dolphin:2016}, adapted from \textit{HST}Phot \citep{Dolphin:2000}. The photometry was run simultaneously on all images using the pipeline snHST \citep{McCully:2018}.

The full set of \emph{HST} photometry of SN~2012Z is presented in \autoref{tab:table1}. The data listed there are in the observer frame; all subsequent uses and presentation of the photometry apply a correction for Milky Way extinction, $E(B-V) = 0.035$ mag \citep{McCully:2014, McCully:2022}. 

\begin{figure*}[t]
    \centering
\includegraphics[width=\textwidth]{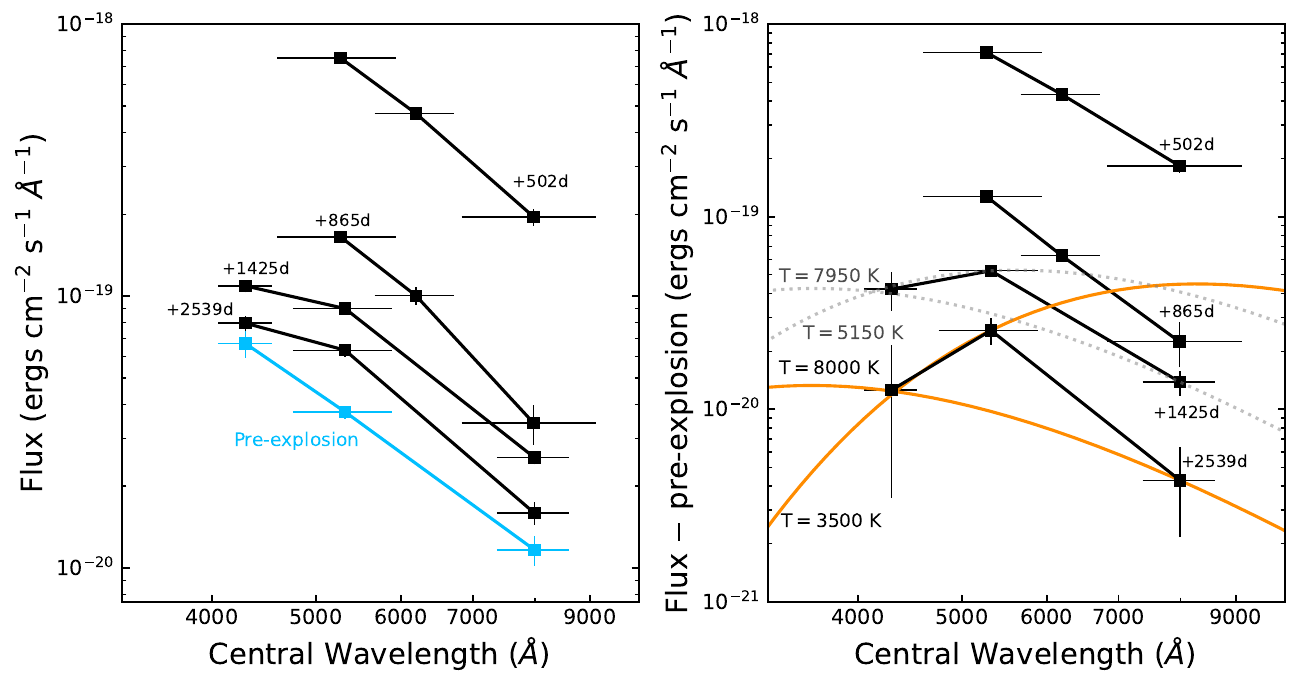}
    \caption{Spectral energy distribution for SN~2012Z. The left panel shows the emission of SN~2012Z in black and the pre-explosion emission in blue. In both panels, the horizontal error bars show the width of the photometric passband. The SN~2012Z SED, at the latest epochs, shows a trend that asymptotes towards the pre-explosion emission, again strongly suggesting the pre-explosion source is still present at the location of the SN at these epochs. The excess F555W emission at +1425 days remains and indeed grows more prominent at +2539 days. This is seen more clearly in the right panel, where the pre-explosion flux is subtracted. No single color temperature can fit the pre-explosion subtracted SED at the latest epochs, but the temperature estimated from the F435W$-$F814W color seems consistent at $\sim$8000 K in the last two epochs. This suggests that it is the more slowly-declining F555W flux that is anomalously bright.} 
    \label{fig:SED}
\end{figure*}

\section{Analysis \label{sec:analysis}}
In this analysis, we adopt a redshift for NGC 1309 of $z~=~0.007125$ \citep{Koribalski:2004} and the Cepheid distance to NGC 1309, $d = 33.0 \pm 1.4$~Mpc \citep{Riess:2011}. We combine the \emph{HST} photometry of SN~2012Z with ground-based data published by \cite{Stritzinger:2015}, and here we adopt their estimates of the date of explosion ($55952.8 \pm 1.5$ MJD) and time of $B$-band maximum ($55967.4 \pm 0.1$ MJD).

\subsection{Late-time Light Curve \label{sec:latelc}}

\autoref{fig:light-curve} shows our 2019 observations of SN~2012Z, extending its light curve by an additional $\sim$1100~days. Notably, even at $\sim$2500 days past maximum light, the source at the location of SN~2012Z is still brighter than pre-explosion, confirming the continued presence of the excess flux in the 2016 observations reported by \cite{McCully:2022}. 

Importantly, we find that the flux is continuing to trend toward the pre-explosion flux in each band, suggesting the source of the pre-explosion flux is still present. It would otherwise need to be a coincidence for the post-explosion light curves to level out close to the pre-explosion magnitudes. \cite{McCully:2022} explore this issue in more detail, examining the idea that the late-time flux should be analyzed independently of the pre-explosion light. In this paper we make the default assumption that in analyzing the late-time photometry, it is most appropriate to subtract the pre-explosion flux. Whether or not the underlying pre-explosion source is the companion helium star as suspected \citep{McCully:2014}, our analysis here focuses on the changes in the system after the supernova.

The decay rates between the last observation at $+$1425~days \citep{McCully:2022} and our most recent observation are represented by dashed lies in the bottom panel in \autoref{fig:light-curve}. The excess, or pre-explosion-subtracted, flux decays at nearly the same rate (just over 1.1 mmag/day) in the F435W and F814W bands while decaying more slowly in F555W, where the excess flux is also most prominent. In \autoref{sec:decay}, we further investigate this extremely late-time decay rate and test whether it can be explained by radioactive decay. If we do not subtract the pre-explosion flux, the decay rate between the last two epochs is much slower, 0.3--0.5 mag/day in all bands.
\\
\\
\subsection{Spectral Energy Distribution and Color Evolution \label{sec:sed}}
At extremely late times, it is difficult to discern between the potential sources that could contribute to the observed excess of flux in \autoref{fig:light-curve}. These potential sources include the companion, possible bound remnant, SN ejecta, CSM interaction, or a light echo. Since spectroscopy is prohibitively expensive at these faint magnitudes, we instead examine the source through its spectral energy distribution (SED). This is a particularly useful tool in uncovering whether or not the pre-explosion flux \citep{McCully:2014} is still present at the location of SN~2012Z. If we assume a scenario in which the pre-explosion flux was dominated by the accretion of the progenitor from the companion, we would expect it to have no contribution to the late-time \textit{HST} photometry observed. However, if we assume that the companion star was the main contributor to the pre-explosion flux, we would expect this flux to contribute to our observations, potentially unchanged after the SN explosion or, through shock-heating, become brighter.

\cite{McCully:2022} found that in the scenario where the pre-explosion flux was no longer present and thus did not impact the late-time photometry, the SED at the last epoch of emission ($+$1425 days) was consistent with a blackbody of temperature 9800 $\pm$ 500 K. On the other hand, if the pre-explosion flux was expected to contribute to the data and thus subtracted, the SED showed a $V$-band excess above a blackbody of temperature 7950~K. \cite{McCully:2022} compared the $V$-band excess at $+$1425 days to models with added [\ion{O}{1}]~6300~\AA\ and \ion{He}{1}~5876~\AA\ line emission but found an unphysical amount of material would be required to produce enough line flux to match the photometry. The absence of an explanation for the F555W excess led \cite{McCully:2022} to slightly favor the scenario without pre-explosion flux contribution to the late-time data.

\autoref{fig:SED} shows the SED of our most recent observation of SN~2012Z ($+$2,359~days) compared to previous epochs, before (left panel) and after (right panel) subtraction of the pre-explosion flux. Similar to the light curve in \autoref{fig:light-curve}, we again see the observations trending toward the pre-explosion flux; for the SED this is most obvious in F435W and F814W (left panel). As described above in \autoref{sec:latelc}, we take this as further evidence that the pre-explosion flux is still present (and relatively unchanged) in the late-time data and focus our analysis on the excess flux (subtracting the pre-explosion light). \cite{McCully:2022} provide a detailed examination of the alternate scenario.

The pre-explosion subtracted data (\autoref{fig:SED}, right panel) shows that F555W or $V$-band excess forms the peak of the SED, as in the previous epoch. The F435W, F555W, and F814W data are not consistent with a single color temperature at either of the latest two epochs ($+$1425 or $+$2539 days). However, the color temperature estimated from F435W and F814W alone is consistent at $\sim$8000 K over these two epochs (spanning more than 1100 days). Thus the light curve and SED data both point to the excess F555W flux as being the anomaly, suggesting an unaccounted-for source of $V$-band flux that is increasing in relative strength. We are not able to resolve this mystery; it may require spectroscopy, e.g., with the next generation of extremely large telescopes. Comparing the flux predicted in F555W by the 8000~K blackbody to the flux observed in F555W (subtracting pre-explosion) at $+$2,359 days, we find the observed flux is 2.7 times higher than predicted. Attempting to model this with a spectral emission line required an implausibly high line peak or width, and so we concur with the conclusion of \cite{McCully:2022} that this green flux excess is unlikely to be the result of a spectral emission line.

\begin{figure}[ht!]
    \centering
\includegraphics[width=\linewidth]{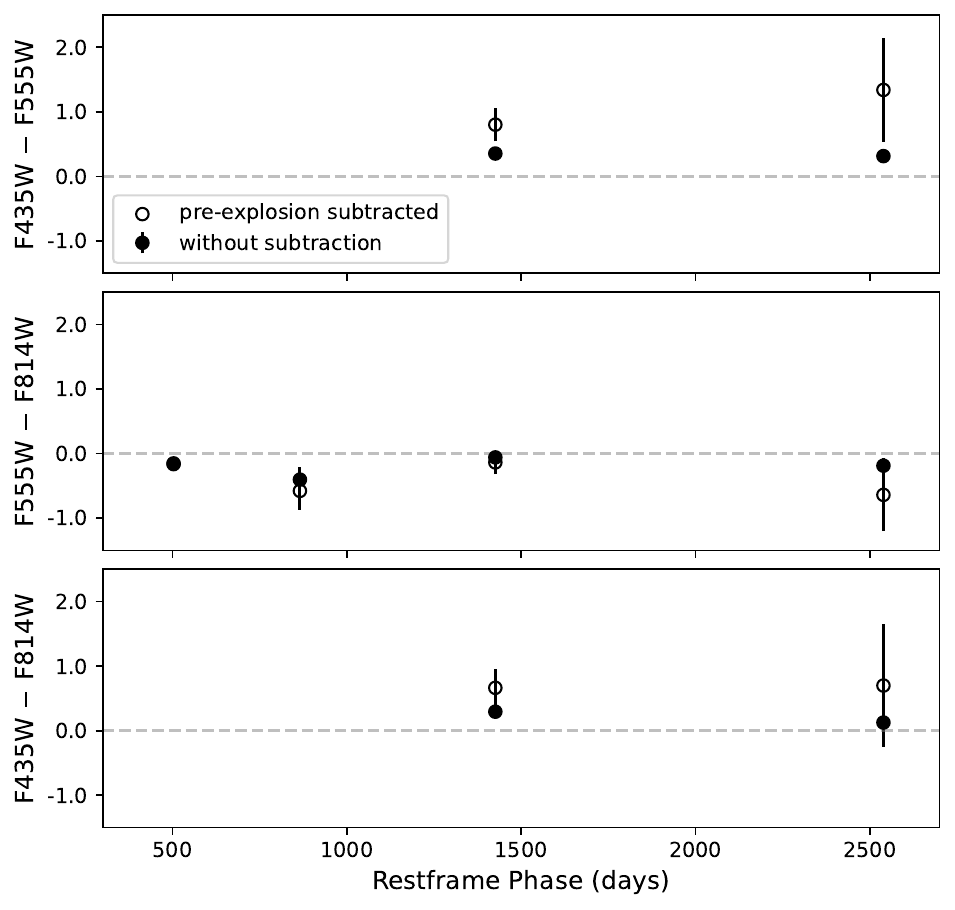}
    \caption{Color evolution of SN~2012Z between 500 and 2500 days post-maximum in  F435W$-$F555W, F555W$-$F814W, and F435W$-$F814W (roughly $B-V$, $V-I$, and $B-I$, respectively), before and after subtraction of the pre-explosion flux. The flux after pre-explosion subtraction maintains a consistent $B-I$ color, while getting redder in $B-V$ and bluer in $V-I$, indicating an anomalous green excess that is becoming more pronounced over time.}
    \label{fig:color_evol}
\end{figure}

The excess flux at the location of SN~2012Z corresponds to a source with a strange SED. \autoref{fig:color_evol} shows the color evolution of SN~2012Z before and after pre-explosion flux subtraction, reiterating the consistent F435W$-$F814W color in the latest two epochs (within the large uncertainties) and increasingly anomalous green emission. At the latest epoch ($+2539$ d), the excess flux has F555W = $27.92 \pm 0.173$ mag, F435W $-$ F555W = 1.34 $\pm$ 0.80 mag and F555W $-$ F814W = $-$0.64 $\pm$ 0.56 mag, and we continue to see the source is notably brighter in F555W than either adjacent filter. 

This SED and color evolution for SN~2012Z disfavor the possibility that the excess flux results from a light echo. Dust scattering light echoes reflect the integrated brightness of the source with a blue scattering function and typically do not show strong color variation \citep{Rest:2012,Graur:2016}. In particular, near maximum light SN~2012Z was relatively blue, with $B-V \approx 0.2$ mag \citep{Stritzinger:2015}, inconsistent with the distinctly green late-time flux excess.

If the binary progenitor system of SN~2012Z ejected circumstellar material (CSM), shock interaction of this material with the supernova ejecta could potentially explain a late-time flux excess \citep{Gerardy:2004, Graham:2019, Dubay:2022, Terwel:2024}. However, CSM interaction models also predict bluer emission than what we observe \citep{Graham:2019,Dubay:2022}, disfavoring this explanation. CSM interaction can also produce optical emission lines that could provide a more promising explanation for the anomalous F555W excess flux. Interaction-powered H$\alpha$ emission has occasionally been detected at late times in normal SNe~Ia \citep[e.g.,][]{Graham:2019,Kollmeier:2019,Vallely:2019,Prieto:2020,Elias-Rosa:2021}, but \cite{Terwel:2024} show this is quite rare, found only in less than 0.5\% of objects. However, H$\alpha$ emission cannot explain the SN~2012Z flux excess we observe: the ACS/WFC F555W passband does not extend redward enough to include H$\alpha$. 

\cite{McCully:2022} investigated two other explanations for the late-time emission, a shock-heated companion star and a radioactively-heated bound remnant. They argued that the previous observation of SN~2012Z at $\sim$1500 days was too red for shock-heated companion models, too blue for radioactive heating, and yet also showed the anomalous green excess. Perhaps surprisingly, our latest epoch measurements of the excess color do not resolve this issue: SN~2012Z did not become bluer or redder in the subsequent 1100 days, just greener!

\subsection{Radioactive Decay \label{sec:decay}} 
\begin{figure*}
    \centering
    \includegraphics[width=\textwidth]{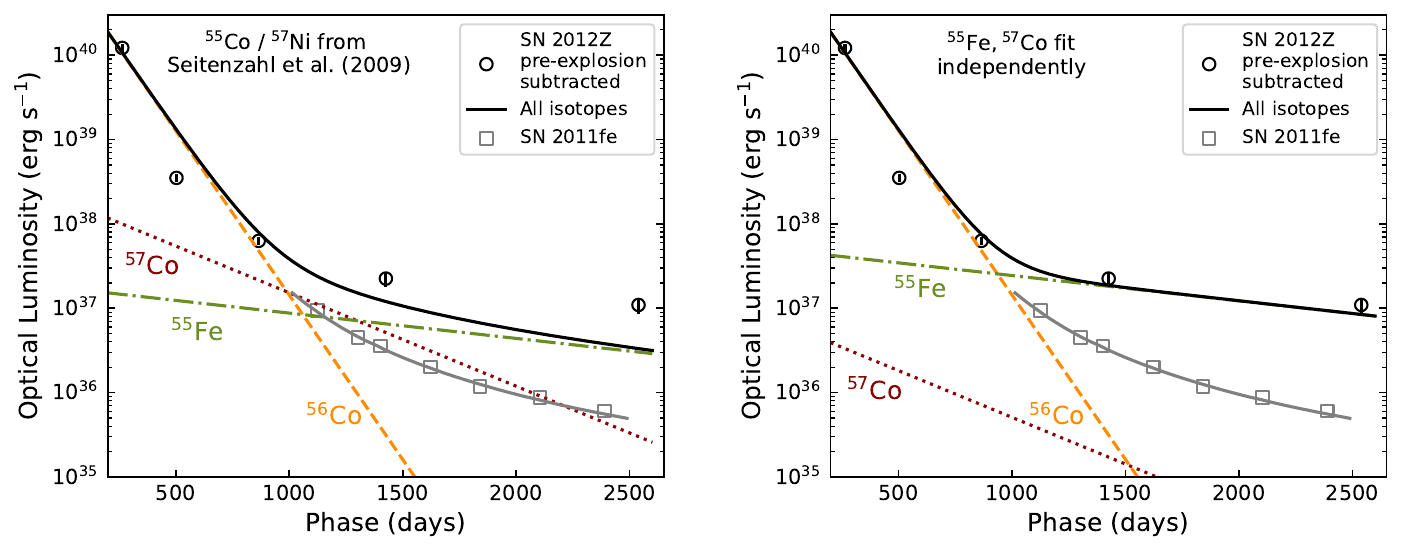}
    \caption{Late-time optical luminosity of SN~2012Z after subtracting the pre-explosion flux (black circles), compared with the normal SN~Ia~2011fe (gray squares; \citealt{Tucker:2022}). SN~2012Z is significantly more luminous than SN 2011fe at similar epochs, even though SN 2011fe is well fit by model predictions of leptonic radioactive decay (gray line). This suggests radioactive decay models for the SN~2012Z excess flux require continued gamma-ray energy deposition, e.g., in a bound remnant. The black solid lines show a combined fit from the decay of $^{56}$Co, $^{55}$Fe, and $^{57}$Co, while the dashed, dash-dotted, and dotted lines show the individual contributions of each of these isotopes, respectively. The two panels show different assumptions about the relative abundances: (\textit{left}) a fit fixing a physically motivated $M_{55}/M_{57}$ ratio from \cite{Seitenzahl:2009}; (\textit{right}) a fit with free $M_{55}$ and $M_{57}$ abundances. Neither of the radioactive fits describe the SN~2012Z light curve well, but the decay rate in the latest two epochs is consistent with $^{55}$Fe.}
    \label{fig:decay}
\end{figure*}

In \autoref{fig:decay} we show the late-time light curve of SN~2012Z spanning $\sim$200 to $\sim$2500 days post-maximum, after subtracting the pre-explosion flux. Plotted along with the data are lines showing the slopes of radioactive decays of isotopes expected to contribute to SN~Ia decay at these epochs. Near peak brightness, the light curve is powered by $^{56}$Ni, after which we expect the decay of $^{56}$Co to power the light curve of normal SNe Ia until approximately a year post-peak brightness \citep{Pankey:1962}. At later times, radioactive isotopes with longer half-lives begin to play important roles in the SN decline, particularly $^{57}$Co ($t_{1/2}\sim272$ days), and even later $^{55}$Fe ($t_{1/2}\sim2.73$ years) \citep{Seitenzahl:2009, Ropke:2012, Graur:2016, Tucker:2022}.

Following \cite{Graur:2016}, we test whether the pre-explosion flux-subtracted late-time observations of SN 2012Z are consistent with the combined radioactive decays of $^{56}$Co, $^{57}$Co, and $^{55}$Fe. Using the \cite{Arnett:1982} law, \cite{Stritzinger:2015} inferred a $^{56}$Ni mass of $\sim 0.3$~M$_{\odot}$ produced by the explosion. We use this value here and fit for the mass ratios $M_{57}/M_{56}$ and $M_{55}/M_{56}$, following the notation of \cite{Dimitriadis:2017} and \cite{Tucker:2022}, i.e. $M_{57}$ is the total mass of all species with mass number 57. The $^{56}$Ni mass derived by \cite{Stritzinger:2015} provides a good fit to the observations at $+$264 and $+$865 days, as expected. The optical luminosity at 502 days, however, is 3.6 times fainter than expected. This unexplained flux deficit may indicate that our radioactive decay model is fatally flawed, but for now we proceed by removing this outlier epoch in the fits. 

In the left panel of \autoref{fig:decay} we first set the ratio of $M_{57}/M_{55}$ to $0.0129/0.0162$, based on the relative abundance of each species to $M_{56}$ in the rpc32 model from \cite{Ohlmann:2014}. We then find a mass ratio of $M_{57}/M_{56} \approx 0.03$ for SN~2012Z ($\chi^2\nu=7.1$), but the best fit is nevertheless 1.9 and 3.3 times fainter than the measured optical luminosities at $+$1426 and $+$2539 days. When the relative mass fractions $M_{57}/M_{56}$ and $M_{55}/M_{56}$ are both fit independently (\autoref{fig:decay}, center panel), we find that the fit is driven by the last two epochs to minimize $M_{57}/M_{56}$ in favor of $M_{55}/M_{56}$, with a resultant $M_{55}/M_{56} \approx 0.1$ ($\chi^2_\nu=3.2)$. This fit produces optical luminosities 1.2 and 1.3 times fainter than the last two observations of SN 2012Z, respectively.

An $M_{55}/M_{56}$ ratio of 0.1 would imply the production of approximately $0.03~M_\odot$ of $^{55}$Co in the explosion, but the model fit requires little to no $^{57}$Ni. This would be quite curious, as we expect both $^{55}$Co and $^{57}$Ni to be produced during the rapid neutronization phase of explosive nucleosynthesis (see, e.g., \citealt{Nomoto:1984}). Moreover, all previously studied models of explosive nucleosynthesis predict a mass fraction ratio
$M_{57}/M_{55}$ greater than unity (see, e.g., \citealt{Tiwari:2022}). It remains to be seen whether simulations of the kind of failed deflagration thought to produce SNe Iax could produce the $^{55}{\rm Co}/^{56}{\rm Ni}$ mass ratio implied by our fits, while at the same time producing very little $^{57}$Ni. One, perhaps implausible, solution could be if somehow the bound remnant preferentially retained $M_{55}$ material compared to $M_{57}$ nuclei.

In \autoref{fig:decay} we also compare the late-time optical luminosity of SN~2012Z to the normal SN~2011fe that has photometry at similarly late phases. \cite{Tucker:2022} estimate the pseudo-bolometric luminosity of SN~2011fe by integrating over the optical SED and adding a constant $35\pm5$\% NIR luminosity. Here, to directly compare with the optical-only luminosity of SN~2012Z, we undo this NIR addition, taking 65\% of the pseudo-bolometric luminosity reported by \cite{Tucker:2022}. They find that the combined leptonic radioactive decay of $^{56}$Co, $^{55}$Fe, and $^{57}$Co, including the effect of delayed deposition \citep{Kushnir&Waxman:2020}, can well explain the late-time light curve of SN~2011fe (solid gray line in \autoref{fig:decay}). By comparison, SN~2012Z is more than 10 times as luminous as SN~2011fe at $+$2500 days, and is declining more slowly. Given that the $^{56}$Ni mass is estimated to be at least 50\% higher for SN~2011fe \citep{Pereira:2013,Mazzali:2015,Dimitriadis:2017,Tucker:2022} than SN~2012Z, and yet SN~2012Z outshines SN~2011fe even at phases earlier than $+$1000 days when $^{56}$Co should still dominate, we are led to the conclusion that if radioactive decay energy explains the late-time light curve of SN~2012Z, more than leptonic decays are required. This may implicate a bound remnant (or its wind), dense enough for gamma-ray energy deposition \citep{Fink:2014,McCully:2014b}.

\subsection{Model Comparisons}

To confront our observations with more physically-motivated models, we investigate the two most likely scenarios for the excess flux, comparing the SN 2012Z flux excess with published bound-remnant wind models and shock-heated helium-star companion models. Following \cite{McCully:2014} and \cite{McCully:2022}, we estimate the optical luminosity of SN~2012Z by interpolating and integrating over the spectral energy distribution from 3400 to 9700~\AA, where our data are best sampled, and estimate uncertainties via bootstrapping. 

\begin{figure}
    \centering
    \includegraphics[width=\linewidth]{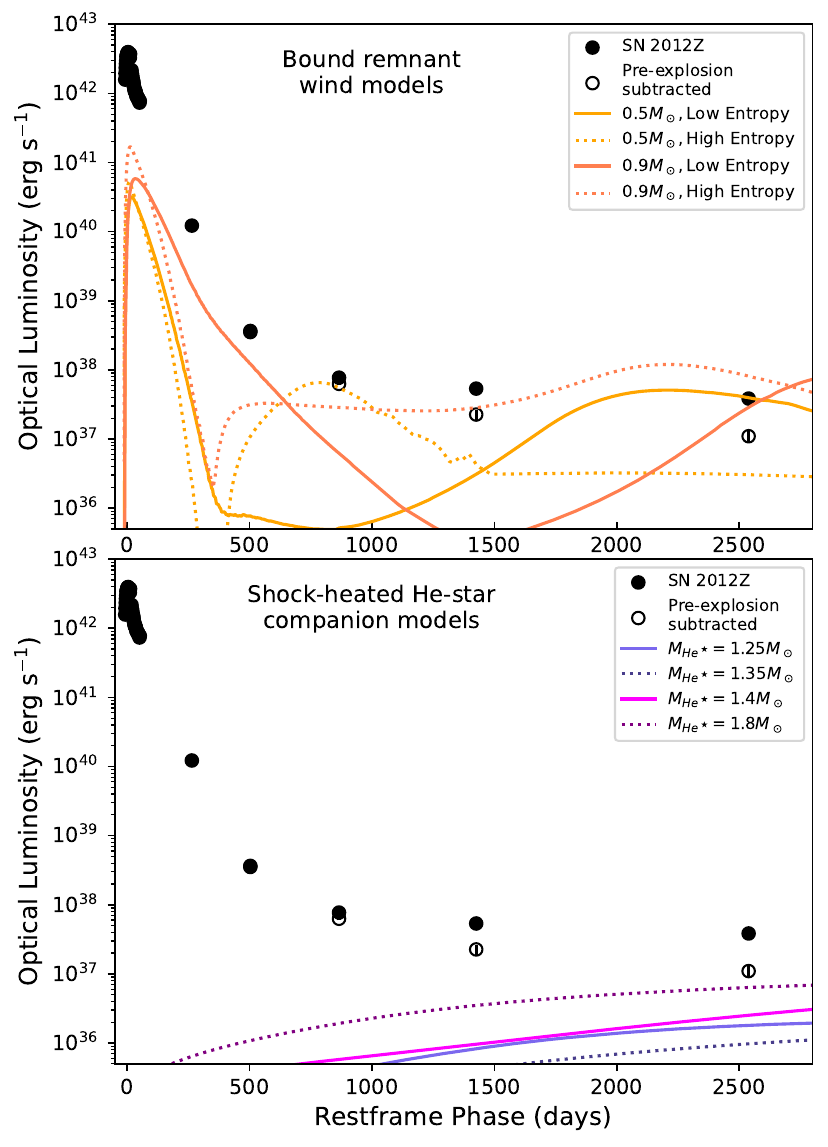}
    \caption{Optical luminosity evolution of SN~2012Z, with early time data from \cite{Stritzinger:2015}, and late time data from \autoref{tab:table1}. \textit{Top:} Comparison of the optical luminosity of SN~2012Z with bound remnant wind models from \cite{ShenSchwab:2017}, labeled by remnant mass and a measure of the initial entropy. The remnant models cool and become very red about a year past maximum where most of their flux drops out of the optical bands. At the latest epochs, the remnant models reheat and produce flux that is comparable to what we see in our latest observation. \textit{Bottom:} Comparison of the optical luminosity of SN~2012Z to the shock-heated helium star companion models from \cite{Pan:2013}. The companion models are much fainter than previous observations out to $\sim1500$~days; however, at our latest epoch, one model for a He star with an initial mass of $1.8~M\odot$ brightens enough to approach the data for SN~2012Z.}
    \label{fig:BR_models}
\end{figure}

In \autoref{fig:BR_models} we compare the optical luminosity of SN~2012Z with the bound-remnant wind models of \cite{ShenSchwab:2017}, using their predicted bolometric luminosity and temperature to synthesize an optical luminosity. In this scenario, the newly produced radioactive material heats the remnant (with initially delayed radioactive decay as electron capture is suppressed in fully ionized material) and drives a radioactively-powered wind. Such a model could explain the low-velocity, high-density (optically thick) permitted-line P-Cygni profiles seen in late-time spectra of SNe Iax \citep{Jha:2006,Foley:2013,McCully:2014b,Camacho-Neves:2023}. All four of the bound-remnant wind models shown in \autoref{fig:BR_models} are bright enough to be comparable to our observations, but none can successfully fit the pre-explosion subtracted luminosity at the latest epochs. Though none of the models are an exact match to the data, they do span a range that is consistent with the observations. We encourage future modeling to determine whether a ``bespoke'' bound-remnant wind model made specifically for SN~2012Z is feasible.

Another possibility to explain the late-time photometric behavior of SNe Iax is increased activity of the donor companion star caused by the impact of the SN ejecta. Pre-explosion imaging suggests the companion star to SN~2012Z is a luminous, blue helium star, and our latest epoch observations indicate that the pre-explosion flux from the companion star is still present. \cite{Pan:2013} model the effects of the explosion of a WD on its He-star companion. They find that the post-impact companion expands and goes through a luminous phase at late times. We compare our data to the post-impact evoution of different helium post-impact remnant stars in the lower panel of \autoref{fig:BR_models}. We find that the companion models are still too faint to compare to the pre-explosion subtracted luminosity even at our latest epoch, implying the excess flux should not be dominated by the flux of the shock-heated companion star. However, the models are beginning to approach our data, re-brightening as light is reprocessed into the optical and the flux of the shock-heated companion increases. This suggests that we are beginning to probe the phases at which these models have the potential to noticeably change the behavior of the luminosity of SN~2012Z, and cause the source to brighten in future observations.

\section{Discussion \& Conclusions \label{sec:conclusions}}

We have presented unprecedented late-time photometry of SN~2012Z, taken with \textit{HST} at 2539 days post maximum light. We find that at the latest epoch the light curve trends toward pre-explosion, implying that the source in the pre-explosion image has not disappeared or changed significantly. \cite{McCully:2022} explores in more detail the scenario in which the pre-explosion flux was dominated by accretion and would no longer contribute to the photometry at these late epochs. Based on our most recent observation, our analysis explores the previously disfavored scenario that the pre-explosion source is still present. \cite{McCully:2014} found the source of pre-explosion to be a luminous blue He-star companion to the progenitor WD that exploded as SN~2012Z. 

We find the continuing presence of a He-star companion to be the most compelling explanation for the continuing source of flux at the level of the pre-explosion flux as naturally explains why the late-time flux is so close to the pre-explosion flux. Alternate models assuming that the flux from pre-explosion, and the late-time flux of SN~2012Z come from totally different mechanisms would need to address the surprising coincidence that the late-time flux appears to trend towards pre-explosion. This is not impossible, given that the bound-remnant models from \cite{ShenSchwab:2017} are independently within a factor of two of the pre-explosion flux; indeed, several mechanisms that we expect to be at play in this regime hover around the Eddington luminosity for a $\sim1~M_\odot$ object. The observed late-time flux may be a composite of multiple sources which happen to have brightnesses similar to those of pre-explosion. 

However, given that the light curve decline rate between the last two epochs (after subtracting pre-explosion) is close to the rate of $^{55}$Fe decay, we favor the interpretation that the pre-explosion flux originates from a He-companion star which continues to be present post-explosion. Although this model appears the most plausible, it does not explain the observed V-band excess, so the system may not be completely unchanged post-explosion. We encourage detailed modeling of this system to investigate whether a combination of accretion and the companion He-star can explain the pre-explosion flux, and then track how the various sources of luminosity (companion star, accretion, SN ejecta, bound-remnant, etc.) contribute post-explosion and how they compare to the late-time data.

Comparison to models of a shock-heated He-star companion from \cite{Pan:2013} show that at $\sim2500$ days post-maximum, we are just beginning to probe the epochs of observation for which a shock-heated companion could significantly influence the light curve of SN~2012Z. At this epoch of observation (as in the previous), the flux of the shock-heated companion is predicted to be too blue to appear in the optical but in later epochs, models suggest that this light could be reprocessed into the optical and the flux from the companion could increase, leading to visible contributions at later phases. Based on these models, even later observations may be able to determine whether or not the companion eventually dominates the light curve.

Subtracting the pre-explosion flux, the late-time behavior of the light curve of SN~2012Z shows similar decay rates in the F435W and F814W filters, while the F555W filter shows a significantly slower decay, appearing as a green flux excess above a blackbody of $8000$~K which fits the other filters well at both $+1425$ and $+2539$~days. This surprising color behavior of SN~2012Z in our latest observation shows the excess flux is too green to be explained by a light echo \citep{Rest:2012,Graur:2016} or CSM interaction \citep{Graham:2019, Dubay:2022}. We instead expect the excess flux at the location of SN~2012Z to be a potential composite of multiple sources: a shock-heated companion, a wind-driven by a bound remnant, and continued radioactive decay in the SN ejecta. 

If the light curve is dominated by radioactive decay at the latest epoch, our analysis suggests that it is dominated by the decay of $^{55}$Fe. In particular, our fits are best when no $^{57}$Co is included. This behavior differs from that of SN~2011fe, which closely follows theoretical predictions for radioactive decay, where $^{57}$Co decay dominates at intermediate times \citep{Tucker:2022}. Despite the decline rate between the last two epochs matching the decay rate of $^{55}$Fe quite well, it would be unphysical to have a large flux contribution from $^{55}$Fe and none from $^{57}$Co. The fact that we do not observe the decay of $^{57}$Co contributing to the late-time light curve of SN~2012Z instead underscores that there are different physical mechanisms at play in SN~2012Z than in SN~2011fe. 

For a normal SN~Ia, radioactive decay models assume an expanding, diluting SN ejecta, for which the gamma-ray opacity becomes negligible at these late times. Thus, the energy deposition is from leptonic processes only (positron decay or electron capture followed by X-ray or Auger electron emission). Here (after pre-explosion subtraction), however, we find that SN~2012Z has an optical luminosity 10--100 times brighter than a normal SN~Ia or predicted by the leptonic decays. Therefore, we argue that the radioactive energy deposition in SN~2012Z is larger than assumed in these models, implying that the gamma-rays must still be trapped in a dense region such as the bound remnant or its wind.  

At our most recent epoch, the latest observation of any SN~Ia to date, the behavior of SN~2012Z adds to the mounting evidence from SNe~Iax in support of a bound remnant. While SNe~Iax share characteristics implying the presence of a bo und remnant such as their relatively slow decay, recently there have been several arguments for direct detections as in the case of the coincident point-source at the location of SN~2008ha \citep{Foley:2014} and the high proper-motion, low mass, Galactic white dwarfs found by \cite{Vennes:2017} and \cite{Raddi:2019} interpreted as partially-burned bound remnants ejected from their systems. With further observations, SN~2012Z may offer the most direct evidence of a bound remnant for this class. 

Models from \cite{ShenSchwab:2017} suggest that a radioactively powered wind from a bound remnant is capable of producing similar luminosities to those we observe at our latest observation of SN~2012Z; however, none of the parameter combinations used in these models match the behavior of SN~2012Z at these late epochs. More comprehensive modeling of bound remnant winds at such late times is required and encouraged to understand the late-time photometric evolution of SN~2012Z and the physical mechanisms responsible.

\vspace{1.5cm}
We thank Ivo Seitenzahl for helpful discussions about late-time behavior of SN~2012Z and possible contribution from radioactive decay. Photometry presented in this paper was obtained by HST program GO-15205 (PI: S.W.~Jha). We gratefully acknowledge support at Rutgers for HST SN~Iax research from NASA/STScI awards HST-GO-15205.001 and HST-GO-16683.001. L.A.K. is supported by a CIERA Postdoctoral Fellowship. C.L. acknowledges support under DOE award DE-SC0010008 to Rutgers University

\textit{Software:} \url{https://github.com/cmccully/snhst} \citep{McCully:2018}, DrizzlePac \citep{2012drzp.book.....G}, Dolphot \citep{Dolphin:2000,Dolphin:2016},  Astropy \citep{Astropy:2013, Astropy:2018}, Matplotlib \citep{Hunter:2007}, NumPy \citep{Harris:2020}, SciPy \citep{Virtanen:2020}, and emcee \citep{Foreman-Mackey:2013}.

\begin{table*}
    \centering
    \setlength{\linewidth}{8pt}
    \caption{\textit{Hubble Space Telescope} Pre-explosion and Late-Time Photometry of SN~2012Z} \label{tab:table1}
    \begin{tabular}{c r r c c c c c c}
    \hline\hline
    Date & Phase & Exposure & Instrument & Detector & Filter & Magnitude & Program & Photometry \\ & (days) &  Time (s) & & & & (mag) & & Reference \\
    \hline
    2005-09-19 & $-$2319 & 9600 & ACS & WFC & F435W & $27.589\pm0.122$ & GO-10497 & \cite{McCully:2014}\\
    2006-03-16 & $-$2142 & 61760 & ACS & WFC & F555W & $27.622\pm0.060$ & GO-10497 & \cite{McCully:2014}\\
    2005-09-01 & $-$2336 & 24000 & ACS & WFC & F814W & $27.532\pm0.135$ & GO-10497 & \cite{McCully:2014}\\
    2010-07-24 & $-$562 & 6991 & WFC3 & IR & F160W & $26.443\pm0.321$ & GO-10711 & \cite{McCully:2014}\\
    2013-06-30 & +502 & 700 & WFC3 & UVIS & F555W & $24.369\pm0.034$ & GO-12913 & \cite{McCully:2022}\\
    2013-06-30 & +502 & 562 & WFC3 & UVIS & F625W & $24.333\pm0.050$ & GO-12913 & \cite{McCully:2022}\\
    2013-06-30 & +502 & 700 & WFC3 & UVIS & F814W & $24.469\pm0.079$ & GO-12913 & \cite{McCully:2022}\\
    2014-06-30 & +865 & 3002 & WFC3 & UVIS & F555W & $26.015\pm0.047$ & GO-13360 & \cite{McCully:2022}\\
    2014-06-30 & +865 & 1952 & WFC3 & UVIS & F625W & $26.009\pm0.079$ & GO-13360 & \cite{McCully:2022}\\
    2014-06-30 & +865 & 2402 & WFC3 & UVIS & F814W & $26.366\pm0.183$ & GO-13360 & \cite{McCully:2022}\\
    2016-01-16 & +1425 & 9624 & ACS &  WFC & F435W & $27.059\pm0.058$ & GO-13757 & \cite{McCully:2022}\\
    2016-01-16 & +1425 & 12642 & ACS & WFC  & F555W & $26.672\pm0.040$ & GO-13757 & \cite{McCully:2022}\\
    2016-01-16 & +1425 & 12868 & ACS & WFC  & F814W & $26.682\pm0.063$ & GO-13757 & \cite{McCully:2022}\\
    2019-02-10 & +2539 & 10016 & ACS & WFC & F435W & $27.402\pm0.070$ & GO-15205 & this paper\\
    2019-02-10 & +2539 & 10632 & ACS & WFC & F555W & $27.055\pm0.060$ & GO-15205 & this paper\\
    2019-02-10 & +2539 & 10704 & ACS & WFC & F814W & $27.193\pm0.104$ & GO-15205 & this paper\\
    \hline
    \end{tabular}
\end{table*}

\clearpage

\normalsize
\bibliography{zotero_abbrev}

\begin{thebibliography}{}
\expandafter\ifx\csname natexlab\endcsname\relax\def\natexlab#1{#1}\fi
\providecommand{\url}[1]{\href{#1}{#1}}
\providecommand{\dodoi}[1]{doi:~\href{http://doi.org/#1}{\nolinkurl{#1}}}
\providecommand{\doeprint}[1]{\href{http://ascl.net/#1}{\nolinkurl{http://ascl.net/#1}}}
\providecommand{\doarXiv}[1]{\href{https://arxiv.org/abs/#1}{\nolinkurl{https://arxiv.org/abs/#1}}}

\bibitem[{{Anderson} \& {Bedin}(2010)}]{AndersonandBendin:2010:}
{Anderson}, J., \& {Bedin}, L. 2010, {An Empirical Pixel-Based Correction for Imperfect CTE. I. HST's Advanced Camera for Surveys}, Instrument Science Report ACS 2010-03

\bibitem[{{Arnett}(1982)}]{Arnett:1982}
{Arnett}, W.~D. 1982, \apj, 253, 785, \dodoi{10.1086/159681}

\bibitem[{{Astropy Collaboration} {et~al.}(2013){Astropy Collaboration}, {Robitaille}, {Tollerud}, {Greenfield}, {Droettboom}, {Bray}, {Aldcroft}, {Davis}, {Ginsburg}, {Price-Whelan}, {Kerzendorf}, {Conley}, {Crighton}, {Barbary}, {Muna}, {Ferguson}, {Grollier}, {Parikh}, {Nair}, {Unther}, {Deil}, {Woillez}, {Conseil}, {Kramer}, {Turner}, {Singer}, {Fox}, {Weaver}, {Zabalza}, {Edwards}, {Azalee Bostroem}, {Burke}, {Casey}, {Crawford}, {Dencheva}, {Ely}, {Jenness}, {Labrie}, {Lim}, {Pierfederici}, {Pontzen}, {Ptak}, {Refsdal}, {Servillat}, \& {Streicher}}]{Astropy:2013}
{Astropy Collaboration}, {Robitaille}, T.~P., {Tollerud}, E.~J., {et~al.} 2013, \aap, 558, A33, \dodoi{10.1051/0004-6361/201322068}

\bibitem[{{Astropy Collaboration} {et~al.}(2018){Astropy Collaboration}, {Price-Whelan}, {Sip{\H{o}}cz}, {G{\"u}nther}, {Lim}, {Crawford}, {Conseil}, {Shupe}, {Craig}, {Dencheva}, {Ginsburg}, {VanderPlas}, {Bradley}, {P{\'e}rez-Su{\'a}rez}, {de Val-Borro}, {Aldcroft}, {Cruz}, {Robitaille}, {Tollerud}, {Ardelean}, {Babej}, {Bach}, {Bachetti}, {Bakanov}, {Bamford}, {Barentsen}, {Barmby}, {Baumbach}, {Berry}, {Biscani}, {Boquien}, {Bostroem}, {Bouma}, {Brammer}, {Bray}, {Breytenbach}, {Buddelmeijer}, {Burke}, {Calderone}, {Cano Rodr{\'\i}guez}, {Cara}, {Cardoso}, {Cheedella}, {Copin}, {Corrales}, {Crichton}, {D'Avella}, {Deil}, {Depagne}, {Dietrich}, {Donath}, {Droettboom}, {Earl}, {Erben}, {Fabbro}, {Ferreira}, {Finethy}, {Fox}, {Garrison}, {Gibbons}, {Goldstein}, {Gommers}, {Greco}, {Greenfield}, {Groener}, {Grollier}, {Hagen}, {Hirst}, {Homeier}, {Horton}, {Hosseinzadeh}, {Hu}, {Hunkeler}, {Ivezi{\'c}}, {Jain}, {Jenness}, {Kanarek}, {Kendrew}, {Kern}, {Kerzendorf}, {Khvalko}, {King}, {Kirkby}, {Kulkarni},
  {Kumar}, {Lee}, {Lenz}, {Littlefair}, {Ma}, {Macleod}, {Mastropietro}, {McCully}, {Montagnac}, {Morris}, {Mueller}, {Mumford}, {Muna}, {Murphy}, {Nelson}, {Nguyen}, {Ninan}, {N{\"o}the}, {Ogaz}, {Oh}, {Parejko}, {Parley}, {Pascual}, {Patil}, {Patil}, {Plunkett}, {Prochaska}, {Rastogi}, {Reddy Janga}, {Sabater}, {Sakurikar}, {Seifert}, {Sherbert}, {Sherwood-Taylor}, {Shih}, {Sick}, {Silbiger}, {Singanamalla}, {Singer}, {Sladen}, {Sooley}, {Sornarajah}, {Streicher}, {Teuben}, {Thomas}, {Tremblay}, {Turner}, {Terr{\'o}n}, {van Kerkwijk}, {de la Vega}, {Watkins}, {Weaver}, {Whitmore}, {Woillez}, {Zabalza}, \& {Astropy Contributors}}]{Astropy:2018}
{Astropy Collaboration}, {Price-Whelan}, A.~M., {Sip{\H{o}}cz}, B.~M., {et~al.} 2018, \aj, 156, 123, \dodoi{10.3847/1538-3881/aabc4f}

\bibitem[{{Camacho-Neves} {et~al.}(2023){Camacho-Neves}, {Jha}, {Barna}, {Dai}, {Filippenko}, {Foley}, {Hosseinzadeh}, {Howell}, {Johansson}, {Kelly}, {Kerzendorf}, {Kwok}, {Larison}, {Magee}, {McCully}, {O'Brien}, {Pan}, {Pandya}, {Singhal}, {Stahl}, {Szalai}, {Wieber}, \& {Williamson}}]{Camacho-Neves:2023}
{Camacho-Neves}, Y., {Jha}, S.~W., {Barna}, B., {et~al.} 2023, \apj, 951, 67, \dodoi{10.3847/1538-4357/acd558}

\bibitem[{{Dimitriadis} {et~al.}(2017){Dimitriadis}, {Sullivan}, {Kerzendorf}, {Ruiter}, {Seitenzahl}, {Taubenberger}, {Doran}, {Gal-Yam}, {Laher}, {Maguire}, {Nugent}, {Ofek}, \& {Surace}}]{Dimitriadis:2017}
{Dimitriadis}, G., {Sullivan}, M., {Kerzendorf}, W., {et~al.} 2017, \mnras, 468, 3798, \dodoi{10.1093/mnras/stx683}

\bibitem[{{Dolphin}(2016)}]{Dolphin:2016}
{Dolphin}, A. 2016, {DOLPHOT: Stellar photometry}, Astrophysics Source Code Library, record ascl:1608.013

\bibitem[{{Dolphin}(2000)}]{Dolphin:2000}
{Dolphin}, A.~E. 2000, \pasp, 112, 1383, \dodoi{10.1086/316630}

\bibitem[{{Dubay} {et~al.}(2022){Dubay}, {Tucker}, {Do}, {Shappee}, \& {Anand}}]{Dubay:2022}
{Dubay}, L.~O., {Tucker}, M.~A., {Do}, A., {Shappee}, B.~J., \& {Anand}, G.~S. 2022, \apj, 926, 98, \dodoi{10.3847/1538-4357/ac3bb4}

\bibitem[{{Elias-Rosa} {et~al.}(2021){Elias-Rosa}, {Chen}, {Benetti}, {Dong}, {Prieto}, {Cappellaro}, {Kollmeier}, {Morrell}, {Piro}, \& {Phillips}}]{Elias-Rosa:2021}
{Elias-Rosa}, N., {Chen}, P., {Benetti}, S., {et~al.} 2021, \aap, 652, A115, \dodoi{10.1051/0004-6361/202141218}

\bibitem[{{Fink} {et~al.}(2014){Fink}, {Kromer}, {Seitenzahl}, {Ciaraldi-Schoolmann}, {R{\"o}pke}, {Sim}, {Pakmor}, {Ruiter}, \& {Hillebrandt}}]{Fink:2014}
{Fink}, M., {Kromer}, M., {Seitenzahl}, I.~R., {et~al.} 2014, \mnras, 438, 1762, \dodoi{10.1093/mnras/stt2315}

\bibitem[{{Foley} {et~al.}(2016){Foley}, {Jha}, {Pan}, {Zheng}, {Bildsten}, {Filippenko}, \& {Kasen}}]{Foley:2016}
{Foley}, R.~J., {Jha}, S.~W., {Pan}, Y.-C., {et~al.} 2016, \mnras, 461, 433, \dodoi{10.1093/mnras/stw1320}

\bibitem[{{Foley} {et~al.}(2014){Foley}, {McCully}, {Jha}, {Bildsten}, {Fong}, {Narayan}, {Rest}, \& {Stritzinger}}]{Foley:2014}
{Foley}, R.~J., {McCully}, C., {Jha}, S.~W., {et~al.} 2014, \apj, 792, 29, \dodoi{10.1088/0004-637X/792/1/29}

\bibitem[{{Foley} {et~al.}(2013){Foley}, {Challis}, {Chornock}, {Ganeshalingam}, {Li}, {Marion}, {Morrell}, {Pignata}, {Stritzinger}, {Silverman}, {Wang}, {Anderson}, {Filippenko}, {Freedman}, {Hamuy}, {Jha}, {Kirshner}, {McCully}, {Persson}, {Phillips}, {Reichart}, \& {Soderberg}}]{Foley:2013}
{Foley}, R.~J., {Challis}, P.~J., {Chornock}, R., {et~al.} 2013, \apj, 767, 57, \dodoi{10.1088/0004-637X/767/1/57}

\bibitem[{{Foreman-Mackey} {et~al.}(2013){Foreman-Mackey}, {Hogg}, {Lang}, \& {Goodman}}]{Foreman-Mackey:2013}
{Foreman-Mackey}, D., {Hogg}, D.~W., {Lang}, D., \& {Goodman}, J. 2013, \pasp, 125, 306, \dodoi{10.1086/670067}

\bibitem[{{Gerardy} {et~al.}(2004){Gerardy}, {H{\"o}flich}, {Fesen}, {Marion}, {Nomoto}, {Quimby}, {Schaefer}, {Wang}, \& {Wheeler}}]{Gerardy:2004}
{Gerardy}, C.~L., {H{\"o}flich}, P., {Fesen}, R.~A., {et~al.} 2004, \apj, 607, 391, \dodoi{10.1086/383488}

\bibitem[{{Gonzaga} {et~al.}(2012){Gonzaga}, {Hack}, {Fruchter}, \& {Mack}}]{2012drzp.book.....G}
{Gonzaga}, S., {Hack}, W., {Fruchter}, A., \& {Mack}, J. 2012, {The DrizzlePac Handbook}

\bibitem[{{Graham} {et~al.}(2019){Graham}, {Harris}, {Nugent}, {Maguire}, {Sullivan}, {Smith}, {Valenti}, {Goobar}, {Fox}, {Shen}, {Kelly}, {McCully}, {Brink}, \& {Filippenko}}]{Graham:2019}
{Graham}, M.~L., {Harris}, C.~E., {Nugent}, P.~E., {et~al.} 2019, \apj, 871, 62, \dodoi{10.3847/1538-4357/aaf41e}

\bibitem[{{Graur} {et~al.}(2016){Graur}, {Zurek}, {Shara}, {Riess}, {Seitenzahl}, \& {Rest}}]{Graur:2016}
{Graur}, O., {Zurek}, D., {Shara}, M.~M., {et~al.} 2016, \apj, 819, 31, \dodoi{10.3847/0004-637X/819/1/31}

\bibitem[{{Harris} {et~al.}(2020){Harris}, {Millman}, {van der Walt}, {Gommers}, {Virtanen}, {Cournapeau}, {Wieser}, {Taylor}, {Berg}, {Smith}, {Kern}, {Picus}, {Hoyer}, {van Kerkwijk}, {Brett}, {Haldane}, {del R{\'\i}o}, {Wiebe}, {Peterson}, {G{\'e}rard-Marchant}, {Sheppard}, {Reddy}, {Weckesser}, {Abbasi}, {Gohlke}, \& {Oliphant}}]{Harris:2020}
{Harris}, C.~R., {Millman}, K.~J., {van der Walt}, S.~J., {et~al.} 2020, \nat, 585, 357, \dodoi{10.1038/s41586-020-2649-2}

\bibitem[{{Hoyle} \& {Fowler}(1960)}]{HoyleandFowler:1960}
{Hoyle}, F., \& {Fowler}, W.~A. 1960, \apj, 132, 565, \dodoi{10.1086/146963}

\bibitem[{{Hunter}(2007)}]{Hunter:2007}
{Hunter}, J.~D. 2007, Computing in Science and Engineering, 9, 90, \dodoi{10.1109/MCSE.2007.55}

\bibitem[{{Jha} {et~al.}(2006){Jha}, {Branch}, {Chornock}, {Foley}, {Li}, {Swift}, {Casebeer}, \& {Filippenko}}]{Jha:2006}
{Jha}, S., {Branch}, D., {Chornock}, R., {et~al.} 2006, \aj, 132, 189, \dodoi{10.1086/504599}

\bibitem[{{Jha}(2017)}]{Jha:2017}
{Jha}, S.~W. 2017, in Handbook of Supernovae, ed. A.~W. {Alsabti} \& P.~{Murdin}, 375, \dodoi{10.1007/978-3-319-21846-5_42}

\bibitem[{{Jha} {et~al.}(2019){Jha}, {Maguire}, \& {Sullivan}}]{Jha:2019}
{Jha}, S.~W., {Maguire}, K., \& {Sullivan}, M. 2019, Nature Astronomy, 3, 706, \dodoi{10.1038/s41550-019-0858-0}

\bibitem[{{Kollmeier} {et~al.}(2019){Kollmeier}, {Chen}, {Dong}, {Morrell}, {Phillips}, {Kushnir}, {Prieto}, {Piro}, \& {Simon}}]{Kollmeier:2019}
{Kollmeier}, J.~A., {Chen}, P., {Dong}, S., {et~al.} 2019, \mnras, 486, 3041, \dodoi{10.1093/mnras/stz953}

\bibitem[{{Koribalski} {et~al.}(2004){Koribalski}, {Staveley-Smith}, {Kilborn}, {Ryder}, {Kraan-Korteweg}, {Ryan-Weber}, {Ekers}, {Jerjen}, {Henning}, {Putman}, {Zwaan}, {de Blok}, {Calabretta}, {Disney}, {Minchin}, {Bhathal}, {Boyce}, {Drinkwater}, {Freeman}, {Gibson}, {Green}, {Haynes}, {Juraszek}, {Kesteven}, {Knezek}, {Mader}, {Marquarding}, {Meyer}, {Mould}, {Oosterloo}, {O'Brien}, {Price}, {Sadler}, {Schr{\"o}der}, {Stewart}, {Stootman}, {Waugh}, {Warren}, {Webster}, \& {Wright}}]{Koribalski:2004}
{Koribalski}, B.~S., {Staveley-Smith}, L., {Kilborn}, V.~A., {et~al.} 2004, \aj, 128, 16, \dodoi{10.1086/421744}

\bibitem[{{Kushnir} \& {Waxman}(2020)}]{Kushnir&Waxman:2020}
{Kushnir}, D., \& {Waxman}, E. 2020, \mnras, 493, 5617, \dodoi{10.1093/mnras/staa690}

\bibitem[{{Li} {et~al.}(2003){Li}, {Filippenko}, {Chornock}, {Berger}, {Berlind}, {Calkins}, {Challis}, {Fassnacht}, {Jha}, {Kirshner}, {Matheson}, {Sargent}, {Simcoe}, {Smith}, \& {Squires}}]{Li:2003}
{Li}, W., {Filippenko}, A.~V., {Chornock}, R., {et~al.} 2003, \pasp, 115, 453, \dodoi{10.1086/374200}

\bibitem[{{Magee} {et~al.}(2016){Magee}, {Kotak}, {Sim}, {Kromer}, {Rabinowitz}, {Smartt}, {Baltay}, {Campbell}, {Chen}, {Fink}, {Gal-Yam}, {Galbany}, {Hillebrandt}, {Inserra}, {Kankare}, {Le Guillou}, {Lyman}, {Maguire}, {Pakmor}, {R{\"o}pke}, {Ruiter}, {Seitenzahl}, {Sullivan}, {Valenti}, \& {Young}}]{Magee:2016}
{Magee}, M.~R., {Kotak}, R., {Sim}, S.~A., {et~al.} 2016, \aap, 589, A89, \dodoi{10.1051/0004-6361/201528036}

\bibitem[{{Mazzali} {et~al.}(2015){Mazzali}, {Sullivan}, {Filippenko}, {Garnavich}, {Clubb}, {Maguire}, {Pan}, {Shappee}, {Silverman}, {Benetti}, {Hachinger}, {Nomoto}, \& {Pian}}]{Mazzali:2015}
{Mazzali}, P.~A., {Sullivan}, M., {Filippenko}, A.~V., {et~al.} 2015, \mnras, 450, 2631, \dodoi{10.1093/mnras/stv761}

\bibitem[{{McCully} {et~al.}(2018){McCully}, {Hosseinzadeh}, {Takaro}, \& {Hiramatsu}}]{McCully:2018}
{McCully}, C., {Hosseinzadeh}, G., {Takaro}, T., \& {Hiramatsu}, D. 2018, {cmccully/snhst: Initial Release}, 0.10.0,  Zenodo, \dodoi{10.5281/zenodo.1482004}

\bibitem[{{McCully} {et~al.}(2014{\natexlab{a}}){McCully}, {Jha}, {Foley}, {Chornock}, {Holtzman}, {Balam}, {Branch}, {Filippenko}, {Frieman}, {Fynbo}, {Galbany}, {Ganeshalingam}, {Garnavich}, {Graham}, {Hsiao}, {Leloudas}, {Leonard}, {Li}, {Riess}, {Sako}, {Schneider}, {Silverman}, {Sollerman}, {Steele}, {Thomas}, {Wheeler}, \& {Zheng}}]{McCully:2014b}
{McCully}, C., {Jha}, S.~W., {Foley}, R.~J., {et~al.} 2014{\natexlab{a}}, \apj, 786, 134, \dodoi{10.1088/0004-637X/786/2/134}

\bibitem[{{McCully} {et~al.}(2014{\natexlab{b}}){McCully}, {Jha}, {Foley}, {Bildsten}, {Fong}, {Kirshner}, {Marion}, {Riess}, \& {Stritzinger}}]{McCully:2014}
---. 2014{\natexlab{b}}, \nat, 512, 54, \dodoi{10.1038/nature13615}

\bibitem[{{McCully} {et~al.}(2022){McCully}, {Jha}, {Scalzo}, {Howell}, {Foley}, {Zeng}, {Liu}, {Hosseinzadeh}, {Bildsten}, {Riess}, {Kirshner}, {Marion}, \& {Camacho-Neves}}]{McCully:2022}
{McCully}, C., {Jha}, S.~W., {Scalzo}, R.~A., {et~al.} 2022, \apj, 925, 138, \dodoi{10.3847/1538-4357/ac3bbd}

\bibitem[{{Miller} {et~al.}(2017){Miller}, {Kasliwal}, {Cao}, {Adams}, {Goobar}, {Kne{\v{z}}evi{\'c}}, {Laher}, {Lunnan}, {Masci}, {Nugent}, {Perley}, {Petrushevska}, {Quimby}, {Rebbapragada}, {Sollerman}, {Taddia}, \& {Kulkarni}}]{Miller:2017}
{Miller}, A.~A., {Kasliwal}, M.~M., {Cao}, Y., {et~al.} 2017, \apj, 848, 59, \dodoi{10.3847/1538-4357/aa8c7e}

\bibitem[{{Nomoto} {et~al.}(1984){Nomoto}, {Thielemann}, \& {Yokoi}}]{Nomoto:1984}
{Nomoto}, K., {Thielemann}, F.~K., \& {Yokoi}, K. 1984, \apj, 286, 644, \dodoi{10.1086/162639}

\bibitem[{{Ohlmann} {et~al.}(2014){Ohlmann}, {Kromer}, {Fink}, {Pakmor}, {Seitenzahl}, {Sim}, \& {R{\"o}pke}}]{Ohlmann:2014}
{Ohlmann}, S.~T., {Kromer}, M., {Fink}, M., {et~al.} 2014, \aap, 572, A57, \dodoi{10.1051/0004-6361/201423924}

\bibitem[{{Pan} {et~al.}(2013){Pan}, {Ricker}, \& {Taam}}]{Pan:2013}
{Pan}, K.-C., {Ricker}, P.~M., \& {Taam}, R.~E. 2013, \apj, 773, 49, \dodoi{10.1088/0004-637X/773/1/49}

\bibitem[{{Pankey}(1962)}]{Pankey:1962}
{Pankey}, Titus, J. 1962, PhD thesis, Howard University, Washington DC

\bibitem[{{Pereira} {et~al.}(2013){Pereira}, {Thomas}, {Aldering}, {Antilogus}, {Baltay}, {Benitez-Herrera}, {Bongard}, {Buton}, {Canto}, {Cellier-Holzem}, {Chen}, {Childress}, {Chotard}, {Copin}, {Fakhouri}, {Fink}, {Fouchez}, {Gangler}, {Guy}, {Hillebrandt}, {Hsiao}, {Kerschhaggl}, {Kowalski}, {Kromer}, {Nordin}, {Nugent}, {Paech}, {Pain}, {P{\'e}contal}, {Perlmutter}, {Rabinowitz}, {Rigault}, {Runge}, {Saunders}, {Smadja}, {Tao}, {Taubenberger}, {Tilquin}, \& {Wu}}]{Pereira:2013}
{Pereira}, R., {Thomas}, R.~C., {Aldering}, G., {et~al.} 2013, \aap, 554, A27, \dodoi{10.1051/0004-6361/201221008}

\bibitem[{{Perley} {et~al.}(2020){Perley}, {Fremling}, {Sollerman}, {Miller}, {Dahiwale}, {Sharma}, {Bellm}, {Biswas}, {Brink}, {Bruch}, {De}, {Dekany}, {Drake}, {Duev}, {Filippenko}, {Gal-Yam}, {Goobar}, {Graham}, {Graham}, {Ho}, {Irani}, {Kasliwal}, {Kim}, {Kulkarni}, {Mahabal}, {Masci}, {Modak}, {Neill}, {Nordin}, {Riddle}, {Soumagnac}, {Strotjohann}, {Schulze}, {Taggart}, {Tzanidakis}, {Walters}, \& {Yan}}]{Perley:2020}
{Perley}, D.~A., {Fremling}, C., {Sollerman}, J., {et~al.} 2020, \apj, 904, 35, \dodoi{10.3847/1538-4357/abbd98}

\bibitem[{{Prieto} {et~al.}(2020){Prieto}, {Chen}, {Dong}, {Bose}, {Gal-Yam}, {Holoien}, {Kollmeier}, {Phillips}, \& {Shappee}}]{Prieto:2020}
{Prieto}, J.~L., {Chen}, P., {Dong}, S., {et~al.} 2020, \apj, 889, 100, \dodoi{10.3847/1538-4357/ab6323}

\bibitem[{{Raddi} {et~al.}(2019){Raddi}, {Hollands}, {Koester}, {Hermes}, {G{\"a}nsicke}, {Heber}, {Shen}, {Townsley}, {Pala}, {Reding}, {Toloza}, {Pelisoli}, {Geier}, {Gentile Fusillo}, {Munari}, \& {Strader}}]{Raddi:2019}
{Raddi}, R., {Hollands}, M.~A., {Koester}, D., {et~al.} 2019, \mnras, 489, 1489, \dodoi{10.1093/mnras/stz1618}

\bibitem[{{Rest} {et~al.}(2012){Rest}, {Sinnott}, \& {Welch}}]{Rest:2012}
{Rest}, A., {Sinnott}, B., \& {Welch}, D.~L. 2012, \pasa, 29, 466, \dodoi{10.1071/AS11058}

\bibitem[{{Riess} {et~al.}(2009{\natexlab{a}}){Riess}, {Macri}, {Li}, {Lampeitl}, {Casertano}, {Ferguson}, {Filippenko}, {Jha}, {Chornock}, {Greenhill}, {Mutchler}, {Ganeshalingham}, \& {Hicken}}]{Riess:2009}
{Riess}, A.~G., {Macri}, L., {Li}, W., {et~al.} 2009{\natexlab{a}}, \apjs, 183, 109, \dodoi{10.1088/0067-0049/183/1/109}

\bibitem[{{Riess} {et~al.}(2009{\natexlab{b}}){Riess}, {Macri}, {Casertano}, {Sosey}, {Lampeitl}, {Ferguson}, {Filippenko}, {Jha}, {Li}, {Chornock}, \& {Sarkar}}]{Riess:2009b}
{Riess}, A.~G., {Macri}, L., {Casertano}, S., {et~al.} 2009{\natexlab{b}}, \apj, 699, 539, \dodoi{10.1088/0004-637X/699/1/539}

\bibitem[{{Riess} {et~al.}(2011){Riess}, {Macri}, {Casertano}, {Lampeitl}, {Ferguson}, {Filippenko}, {Jha}, {Li}, \& {Chornock}}]{Riess:2011}
---. 2011, \apj, 730, 119, \dodoi{10.1088/0004-637X/730/2/119}

\bibitem[{{R{\"o}pke} {et~al.}(2012){R{\"o}pke}, {Kromer}, {Seitenzahl}, {Pakmor}, {Sim}, {Taubenberger}, {Ciaraldi-Schoolmann}, {Hillebrandt}, {Aldering}, {Antilogus}, {Baltay}, {Benitez-Herrera}, {Bongard}, {Buton}, {Canto}, {Cellier-Holzem}, {Childress}, {Chotard}, {Copin}, {Fakhouri}, {Fink}, {Fouchez}, {Gangler}, {Guy}, {Hachinger}, {Hsiao}, {Chen}, {Kerschhaggl}, {Kowalski}, {Nugent}, {Paech}, {Pain}, {Pecontal}, {Pereira}, {Perlmutter}, {Rabinowitz}, {Rigault}, {Runge}, {Saunders}, {Smadja}, {Suzuki}, {Tao}, {Thomas}, {Tilquin}, \& {Wu}}]{Ropke:2012}
{R{\"o}pke}, F.~K., {Kromer}, M., {Seitenzahl}, I.~R., {et~al.} 2012, \apjl, 750, L19, \dodoi{10.1088/2041-8205/750/1/L19}

\bibitem[{{Sahu}(2021)}]{2021wfcd.book....5S}
{Sahu}, K. 2021, in WFC3 Data Handbook v. 5, Vol.~5, 5

\bibitem[{{Seitenzahl} {et~al.}(2009){Seitenzahl}, {Taubenberger}, \& {Sim}}]{Seitenzahl:2009}
{Seitenzahl}, I.~R., {Taubenberger}, S., \& {Sim}, S.~A. 2009, \mnras, 400, 531, \dodoi{10.1111/j.1365-2966.2009.15478.x}

\bibitem[{{Shen} \& {Schwab}(2017)}]{ShenSchwab:2017}
{Shen}, K.~J., \& {Schwab}, J. 2017, \apj, 834, 180, \dodoi{10.3847/1538-4357/834/2/180}

\bibitem[{{Srivastav} {et~al.}(2022){Srivastav}, {Smartt}, {Huber}, {Chambers}, {Angus}, {Chen}, {Callan}, {Gillanders}, {McBrien}, {Sim}, {Fulton}, {Hjorth}, {Smith}, {Young}, {Auchettl}, {Anderson}, {Pignata}, {de Boer}, {Lin}, \& {Magnier}}]{Srivastav:2022}
{Srivastav}, S., {Smartt}, S.~J., {Huber}, M.~E., {et~al.} 2022, \mnras, 511, 2708, \dodoi{10.1093/mnras/stac177}

\bibitem[{{Stritzinger} {et~al.}(2015){Stritzinger}, {Valenti}, {Hoeflich}, {Baron}, {Phillips}, {Taddia}, {Foley}, {Hsiao}, {Jha}, {McCully}, {Pandya}, {Simon}, {Benetti}, {Brown}, {Burns}, {Campillay}, {Contreras}, {F{\"o}rster}, {Holmbo}, {Marion}, {Morrell}, \& {Pignata}}]{Stritzinger:2015}
{Stritzinger}, M.~D., {Valenti}, S., {Hoeflich}, P., {et~al.} 2015, \aap, 573, A2, \dodoi{10.1051/0004-6361/201424168}

\bibitem[{{Taubenberger}(2017)}]{Taubenberger:2017}
{Taubenberger}, S. 2017, in Handbook of Supernovae, ed. A.~W. {Alsabti} \& P.~{Murdin}, 317, \dodoi{10.1007/978-3-319-21846-5_37}

\bibitem[{{Terwel} {et~al.}(2024){Terwel}, {Maguire}, {Dimitriadis}, {Smith}, {Reusch}, {Lacroix}, {Galbany}, {Burgaz}, {Harvey}, {Schulze}, {Rigault}, {Groom}, {Hale}, {Kasliwal}, {Kim}, {Purdum}, {Rusholme}, {Sollerman}, {Anderson}, {Chen}, {Frohmaier}, {Gromadzki}, {M{\"u}ller-Bravo}, {Nicholl}, {Srivastav}, \& {Deckers}}]{Terwel:2024}
{Terwel}, J.~H., {Maguire}, K., {Dimitriadis}, G., {et~al.} 2024, arXiv e-prints, arXiv:2402.16962, \dodoi{10.48550/arXiv.2402.16962}

\bibitem[{{Tiwari} {et~al.}(2022){Tiwari}, {Graur}, {Fisher}, {Seitenzahl}, {Leung}, {Nomoto}, {Perets}, \& {Shen}}]{Tiwari:2022}
{Tiwari}, V., {Graur}, O., {Fisher}, R., {et~al.} 2022, \mnras, 515, 3703, \dodoi{10.1093/mnras/stac1618}

\bibitem[{{Tucker} {et~al.}(2022){Tucker}, {Shappee}, {Kochanek}, {Stanek}, {Ashall}, {Anand}, \& {Garnavich}}]{Tucker:2022}
{Tucker}, M.~A., {Shappee}, B.~J., {Kochanek}, C.~S., {et~al.} 2022, \mnras, 517, 4119, \dodoi{10.1093/mnras/stac2873}

\bibitem[{{Vallely} {et~al.}(2019){Vallely}, {Fausnaugh}, {Jha}, {Tucker}, {Eweis}, {Shappee}, {Kochanek}, {Stanek}, {Chen}, {Dong}, {Prieto}, {Sukhbold}, {Thompson}, {Brimacombe}, {Stritzinger}, {Holoien}, {Buckley}, {Gromadzki}, \& {Bose}}]{Vallely:2019}
{Vallely}, P.~J., {Fausnaugh}, M., {Jha}, S.~W., {et~al.} 2019, \mnras, 487, 2372, \dodoi{10.1093/mnras/stz1445}

\bibitem[{{Vennes} {et~al.}(2017){Vennes}, {Nemeth}, {Kawka}, {Thorstensen}, {Khalack}, {Ferrario}, \& {Alper}}]{Vennes:2017}
{Vennes}, S., {Nemeth}, P., {Kawka}, A., {et~al.} 2017, Science, 357, 680, \dodoi{10.1126/science.aam8378}

\bibitem[{{Virtanen} {et~al.}(2020){Virtanen}, {Gommers}, {Oliphant}, {Haberland}, {Reddy}, {Cournapeau}, {Burovski}, {Peterson}, {Weckesser}, {Bright}, {van der Walt}, {Brett}, {Wilson}, {Millman}, {Mayorov}, {Nelson}, {Jones}, {Kern}, {Larson}, {Carey}, {Polat}, {Feng}, {Moore}, {VanderPlas}, {Laxalde}, {Perktold}, {Cimrman}, {Henriksen}, {Quintero}, {Harris}, {Archibald}, {Ribeiro}, {Pedregosa}, {van Mulbregt}, \& {SciPy 1. 0 Contributors}}]{Virtanen:2020}
{Virtanen}, P., {Gommers}, R., {Oliphant}, T.~E., {et~al.} 2020, Nature Methods, 17, 261, \dodoi{10.1038/s41592-019-0686-2}

\bibitem[{{Wang} \& {Han}(2012)}]{Wang:2012}
{Wang}, B., \& {Han}, Z. 2012, \nar, 56, 122, \dodoi{10.1016/j.newar.2012.04.001}

\end{thebibliography}

\end{document}